# Polaritonic materials fabricated and tested with ultrashort-pulse lasers

David W. Ward, Eric Statz, Thomas Feurer, and Keith A. Nelson
The Massachusetts Institute of Technology,
Cambridge, MA 02139, USA

## ABSTRACT

Using femtosecond laser machining, we have fabricated photonic bandgap materials that influence propagation of phonon-polaritons in ferroelectric crystals. Broadband polaritons were generated with impulsive stimulated Raman scattering (ISRS) using an ultrashort laser pulse, and the spatial and temporal evolution of the polaritons were imaged as they propagated through the fabricated structures with polariton real-space imaging. These techniques offer a new approach to optical materials design.

## INTRODUCTION

Polaritonics constitutes an intermediate regime between electronics, with frequencies typically less than 100 GHz, and photonics, with frequencies on the order of 100's of THz. In the former, radiation is propagated by strong coupling to the motion of free electrons in a transmission line, and in the latter radiation is coupled only weakly to bound electronic motion, if at all, and resembles a freely propagating electromagnetic wave with speeds on the order of light in vacuum. Polaritonics occupies the band of frequencies between roughly 100 GHz and 10 THz and relies on elementary excitations known as phonon-polaritons in which the radiation field is strongly coupled to bound ionic charge motion associated with polar transverse optic (TO) phonon modes (*1, 2*). This coupling gives rise to a splitting between the TO and longitudinal optic (LO) phonon frequencies, between which electromagnetic propagation is forbidden, and upper and lower polariton branches whose dispersion properties are well known (3, 4).

When the crystal is patterned with periodic 'air' holes, the dispersive properties are modified such that in addition to the polariton bandgap, a photonic bandgap related to the dielectric periodicity is introduced. Recently the interplay between the intrinsic LO-TO bandgap and a photonic bandgap in an overlapping frequency range that is produced through fabrication of a periodic structure in the crystalline host has been discussed (5, 6). Here we demonstrate the capability for fabrication of polaritonic bandgap structures in the THz range by machining 'air' holes into an $MgO:LiNbO_3$ crystal, generating coherent THz phonon-polaritons in the material through impulsive stimulated Raman scattering (ISRS) (7), and monitoring THz wave propagation through the photonic bandgap materials (PBM's) with polariton imaging (8).

## EXPERIMENTAL DETAILS

A homebuilt multi-pass amplified Ti:Saphire femtosecond amplifier (800 nm, 50 fs FWHM, 1 KHz rep. rate, 700 µJ/pulse) seeded by a KM Labs oscillator (790 nm, 15 fs, 88 MHz rep. rate, 3 nj/pulse) was used in both fabrication and imaging of patterned materials.

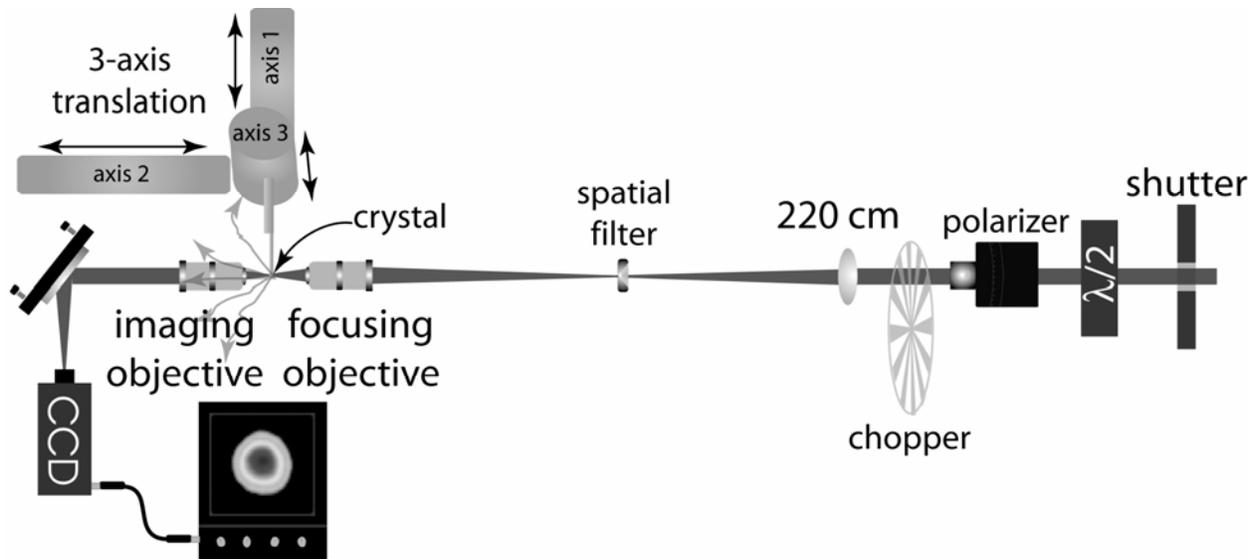

**Figure 1:** Femtosecond laser machining setup. User specified patterns are milled into MgO:LiNbO$_3$ by moving the sample through the beam focus of the microscope objective. Additional components are for control of laser exposure time, pulse energy, and an imaging system for crystal alignment to the beam focus (an experimentally measured beam profile is shown here).

Patterning was carried out in 11 x 10 x 0.034 mm poled, stoichiometric, X-cut MgO:LiNbO$_3$ crystals by focusing amplified femtosecond laser pulses into the crystal with a microscope objective (NA 0.1). This is discussed in detail below and has been described previously for fabrication of guided wave structures (9-11).

Broadband polaritons were created by focusing a femtosecond laser excitation pulse, polarized along the optic axis, into a ferroelectric crystal with a cylindrical lens, resulting in a `line excitation' ripple that propagates laterally away from the excitation region in both directions. The center frequency is inversely proportional to the excitation spot size and is easily determined by calculation or simulation (12). The polariton wavelength was greater than the crystal thickness, such that propagation was predominantly in the lowest order TE waveguide mode of the crystal (13, 14).

As polaritons propagate through the crystal, they modulate the index of refraction, creating a spatio-temporal phase pattern that corresponds to the polariton amplitude distribution over the spatial extent of the crystal. Polariton imaging employs well known techniques of phase-to-amplitude conversion to image this phase pattern onto a CCD camera (15). An image is recorded each time the probe delay is stepped forward with respect to the excitation pulse. The complete set of sequential images acquired during an experiment constitutes a 'movie' that captures the propagating THz wave as a function of time (8).

**FABRICATION OF PHOTONIC BANDGAP MATERIALS WITH ULTRASHORT LASER MACHINING**

The advantages of femtosecond laser machining for a wide range of materials including wide-bandgap dielectrics like like MgO:LiNbO$_3$ have been well documented (16). The experimental setup used for fabrication is illustrated in Fig. 1. It consists of a mechanical shutter to control the number of pulses to which the crystal is exposed; polarization selective optics to control the beam intensity; a microscope focusing objective (NA 0.1); a spatial filter; a phase locked chopper to control repetition rate; and a computer actuated, three-axis translation stage to move the crystal through the focus of the beam. A user specified pattern encoded in HPGL is loaded into a Labview program that interfaces with the translation stage and beam shutter. Upon execution, the program directs the crystal position and exposure time such that the input pattern is reproduced in the crystal with damage extending throughout the crystal thickness.

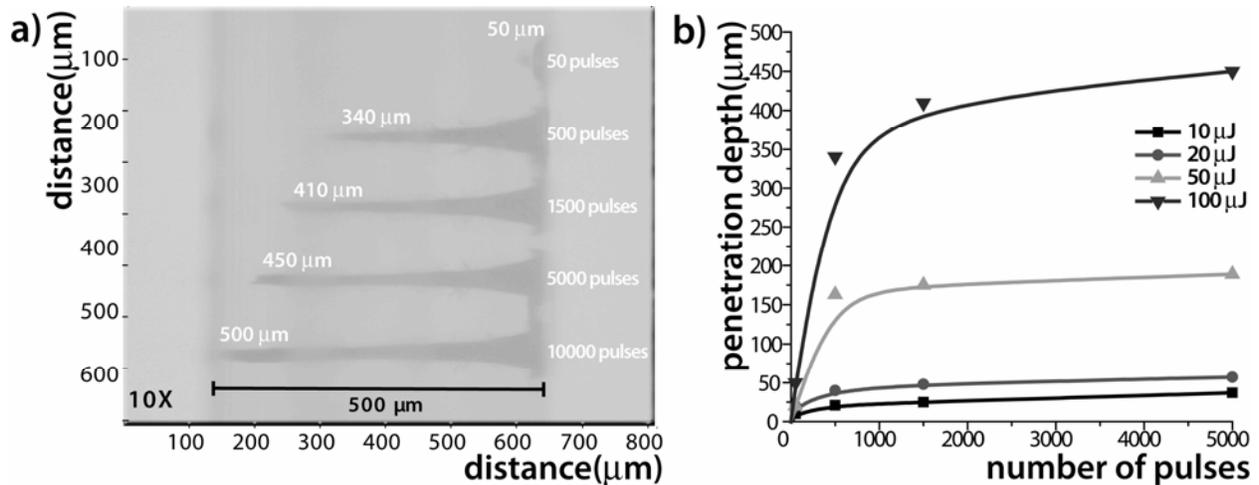

**Figure 2:** Laser induced damage assays. In order to fabricate arbitrary structures in MgO:LiNbO$_3$, details of the laser induced damage were discerned by collecting data on the extent of damage with pulse energy and exposure time. a) Optical micrograph (10X) of a cross-section through the crystal thickness showing laser damage penetration depth with 100 μJ pulses. b) Representation of several data sets showing variation in penetration depth with pulse energy and exposure time (number of pulses) and a fit to form $a\left(1 - \exp(-b \cdot x)\right)$, where $a$ is the asymptote.

The machining parameters are largely dependent on the desired structure, but mainly on the crystal thickness. PBM's and other structures consisting of periodic `air' holes, in which a cylindrical column of material is removed throughout the crystal thickness, are produced by exposing the crystal to the laser beam for a sufficient interval to achieve the desired hole radius and penetration depth, translating by the periodicity of the structure with the shutter closed, and repeating at the new position in the crystal. The PBM's were fabricated with thirty 20 μJ pulses for each 'air' hole in 34 micron thick, poled, stoichiometric, x-cut MgO:LiNbO$_3$. Both the damage radius and penetration depth increase monotonically with exposure time toward asymptotic values. Variation in hole radius is observed between 2-500 pulses and is limited by the beam waist at the focus, while penetration depth depends on the pulse energy (see Fig. 2) and is ultimately limited by the Raleigh range. The pulse fluences employed here greatly exceed the damage threshold in MgO:LiNbO$_3$ in order to achieve damage that extends throughout the thickness of the crystal. Further, through-plane damage is observed to have a slight bevel that tapers off to a constant diameter in the first 300 microns from the crystal surface. This is likely due to wave guiding from a combination of the previously machined channel's dielectric contrast and self-focusing. A sacrificial layer of material may be employed to obtain uniform through-plane damage.

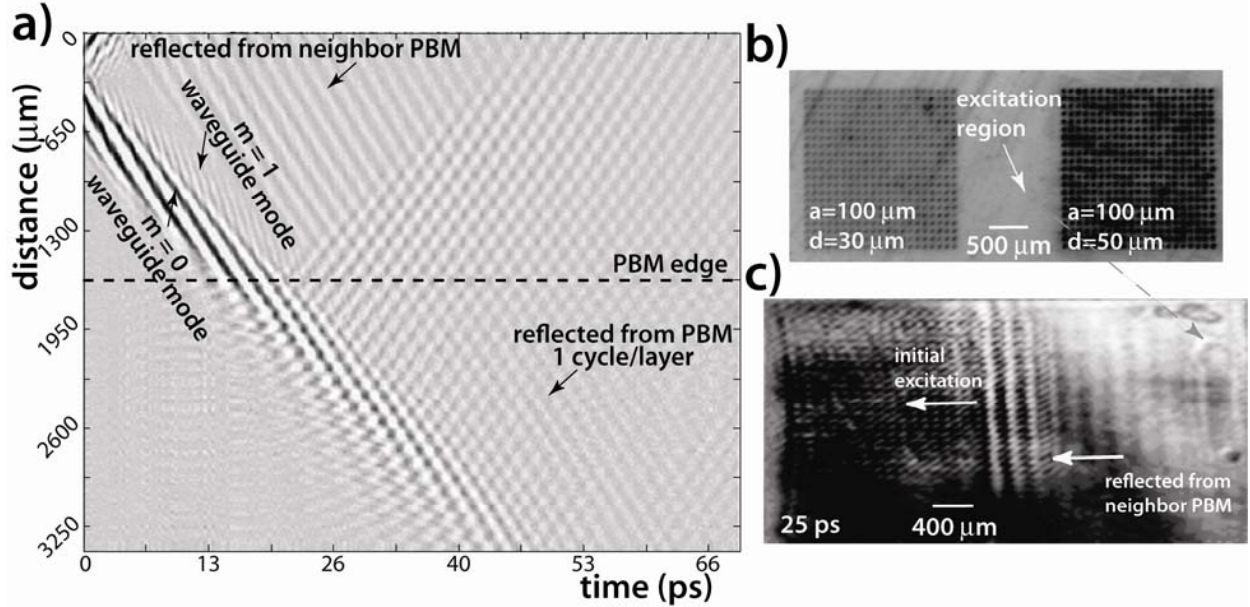

**Figure 3**: Square lattice photonic bandgap material. A broadband THz 'line' excitation is generated in the region indicated in b). An image of the THz fields 25 ps after excitation is shown in c). The initial left-going waveform is shown propagating through the left most structure in b). Also shown is the reflected radiation from the right-going THz waveform, which corresponds to frequency components in the bandgap of the right most structure. a) Space-time plot constructed by integrating along the vertical axis in c) for each frame recorded at different probe-delays. Reflections from both PBM structures are evident as are the two slab waveguide modes of the crystal slab (indicated in figure). These waveguide modes are due to confinement in the dimension of the crystal thickness. The periodicity, *a*, and hole diameter, *d*, are indicated in b).

## TESTING PHOTONIC BANDGAP MATERIALS WITH POLARITON IMAGING

A photonic bandgap is achieved when electromagnetic radiation traverses a spatially periodic dielectric (17, 18). For a 2D array of holes with periodicity *a*, we expect the bandgap to be centered at wavelengths in the vicinity of 2*a*. Radiation with frequencies in this bandgap incident upon the PBM will undergo reflection, with little to no propagation within the PBM. The bandgap is a result of an avoided crossing between modes predominantly in the air and predominantly in the crystal which without the presence of the other would normally have the same frequency. The lower end of the bandgap corresponds to the mode predominantly in the crystal since the crystal has a higher index of refraction, and the upper end of the bandgap corresponds to the mode predominantly in the 'air' holes. Since polariton imaging relies on the ferroelectric host material for the signal, it is not possible to image the electric fields in the 'air' holes; however, the full bandgap will be evident in the linewidth of reflection measurements. Optimal bandgaps in a photonic bandgap slab are achieved by choosing the slab thickness to be on the order of half the wavelength in the dielectric (18). The slab dispersive properties will differ from those in bulk since some fraction of the in-plane electromagnetic radiation is in air, giving rise to a slightly lower effective index of refraction (13, 14). The presence of a PBM in a dielectric slab introduces an additional deviation from bulk dispersion because the 'air' holes further reduce the effective index of refraction.

Figure 3 illustrates the visualization of THz waveform propagation in a square lattice PBM. Broadband, nearly single cycle polaritons were generated in between two PBM's (Fig. 3b,c) by a

'line excitation' of light, cylindrically focused to a width of approximately 130 microns, whose length exceeded the extent of the PBM's in the vertical dimension of Fig. 3b. Their spatiotemporal evolution was recorded through images like that shown in Fig. 3c. Since the polariton responses within the PBM's showed no systematic variation in the vertical dimension, a space-time plot (Fig. 3a) was constructed by compression of the images in this dimension. The compressed images display only one spatial dimension (horizontal in Fig. 3b, vertical in Fig. 3a) vs. probe pulse delay time. The reflected light in the bandgap from both PBM's is clearly visible. It is also apparent that the reflected light has a much narrower linewidth than the incident. The incident broadband waveform has a center frequency of approximately 290 GHz with 400 GHz bandwidth FWHM. The reflected waveform has a frequency of 250 GHz with less than 10 GHz of bandwidth FWHM and a wavelength of approximately 200 microns, which corresponds to 2*a*. Also shown (Fig. 3c) is a single frame, 25 ps after polariton excitation, which illustrates propagation through the PBM. Visualization inside the PBM may not be possible for larger filling fractions, as real-space imaging is only possible in the ferroelectric portion of the PBM.

## CONCLUSIONS

We have demonstrated the femtosecond laser fabrication and testing of polaritonic bandgap structures in ferroelectric crystals. PBM's in this frequency range may find a wide range of applications in integrated THz structures. Because of the capability for direct visualization of the electric fields that are propagated through and reflected from polaritonic bandgap structures, these techniques—ferroelectric PBM's and polariton imaging—constitute a novel procedure for design and implementation of PBM's at any wavelength due to the scalable properties of Maxwell's equations. Further, the propagation time of radiation through the structures, which is analogous to the computation time in a simulation or calculation, makes this procedure highly competitive with existing computer techniques for PBM design. We note that improvements in real-space image quality are possible with higher spatial resolution and with the use of phase-to-amplitude conversion methods that preserve optimal focusing of the polaritons and the fabricated structures.

## ACKNOWLEDGEMENTS

This work was supported in part by the National Science Foundation (CHE-0212375 and MRSEC Program, Grant No. DMR-0213282) and by the Cambridge-MIT Institute grant no. CMI-001.